\documentclass[12pt]{article}

\usepackage{amsfonts,amsmath,amssymb,epsf,bbm,color,array,colortbl,xypic}
\usepackage[english]{babel}
\usepackage{ifpdf}
\ifpdf
\usepackage{epstopdf,hyperref}
\else
\usepackage[hypertex]{hyperref}
\fi

\advance\textheight by \topskip
\usepackage{rotating}

\usepackage{amsthm} 
\newtheorem{Thm}{Theorem}

\theoremstyle{definition}

\newtheorem{Def}[Thm]{Definition}
\numberwithin{equation}{section}
\numberwithin{Thm}{section}

\newcommand\void[1]       {}

\newcommand\be            {\begin{equation}}
\newcommand\bea           {\begin{equation}\begin{array}l\displaystyle}
\newcommand\bearll        {\begin{array}{ll}\displaystyle}
\newcommand\ee            {\end{equation}}
\newcommand\eear          {\end{array}}

\renewcommand{\epsilon}{\varepsilon}

\newcommand\arxiv[2]      {\href{http://arXiv.org/abs/#1}{\tt #2}}
\newcommand\doi[2]        {\href{http://dx.doi.org/#1}{#2}}

\begin{document}
\thispagestyle{empty}
\def\thefootnote{\fnsymbol{footnote}}
\vskip 2.5em
\begin{center}\LARGE
Fusion Rules of\\
the ${\cal W}_{p,q}$ Triplet Models
\end{center}\vskip 2em
\begin{center}\large
  Simon Wood%
  $^{a}$\footnote{Email: {\tt swood@itp.phys.ethz.ch}}
\end{center}
\begin{center}\it$^a$
Institute for Theoretical Physics, ETH Zurich \\
8093 Z\"urich, Switzerland
\end{center}
\vskip 1em
\begin{center}
July 2009
\end{center}
\vskip 1em
\begin{abstract}
In this paper we determine the fusion rules of the logarithmic \(\mathcal{W}_{p,q}\) triplet theory
and construct the Grothendieck group with subgroups for
which consistent product structures can
be defined. The fusion rules are then used to determine
projective covers. This allows us also to write down a candidate for a modular invariant partition function.
Our results demonstrate that recent work on the \(\mathcal{W}_{2,3}\) model 
generalises naturally to arbitrary \((p,q)\).
\end{abstract}

\setcounter{footnote}{0}
\def\thefootnote{\arabic{footnote}}

\newpage

\tableofcontents

\newpage

\section{Introduction}

Logarithmic conformal field theories appear in the description of critical points in many interesting physical
systems. Some examples are polymers, spin chains, percolation, and sand-pile models, see for example 
\cite{Jeng:2006tg,Pearce:2006we,Read:2007qq,Ruelle:2007kg,Mathieu:2007pe,Rasmussen:2008ii,Ridout:2008cv,SaintAubin:2008,Nigro:2009si}
for some recent papers. 
A lot of effort has been invested recently to try and understand these theories in a general context.
For example the logarithmic conformal field theories from the \((1,p)\)-series have been studied
in quite some detail and their structure is now largely understood
\cite{Gaberdiel:1996np,Fuchs:2003yu,Carqueville:2005nu,Gaberdiel:2007jv,Adamovic:2007er,nagatomo:2009,GabKau96a}.
However the more general \((p,q)\)-series for \(p,q\) coprime and \(p,q\geq2\) are not as well understood yet,
though there has been some progress 
recently \cite{GabRunWoo,Rasmussen:2008xi,Rasmussen:2008ez,Feigin:2006iv}.
These theories are referred to as the \(\mathcal{W}_{p,q}\) triplet models, and they can be naturally associated to the minimal models for \(p,q\geq2\).

The goal of this paper is to generalise the results of \cite{GabRunWoo}, where the \(\mathcal{W}_{2,3}\) model was
studied, to general \((p,q)\). In particular we obtain the fusion rules
of the \(\mathcal{W}_{p,q}\) triplet models. This is obviously a prerequisite for any detailed analysis of this theory. We also study
the Grothendieck group that plays a vital role in the boundary description of conformal field theories.
One novel feature of the \(\mathcal{W}_{p,q}\) models is that
the vacuum representation is reducible but indecomposable and we believe that it is responsible
for the fact that the structure of the \(\mathcal{W}_{p,q}\) models, though closely related to that of the minimal models 
and the \(\mathcal{W}_{1,p}\) models, is a lot more complicated than either.

We determine the fusion rules by first generalising the representations appearing in \cite{GabRunWoo} for arbitrary
\(p,q\) in Section \ref{sec:repcont}. For a subset of these representations the fusion rules have already been
determined in \cite{Rasmussen:2008xi,Rasmussen:2008ez} and we will propose a way to extend these rules in an associative
manner to all the other representations in section \ref{sec:fusionrules} (with consistency checks in appendix 
\ref{sec:consistency} for certain explicit values of \((p,q)\)).
In section \ref{sec:grothendieckgroup} we determine the subgroup
of the Grothendieck group for which a consistent product structure induced by the fusion rules can be defined. In section
\ref{sec:projective} we address the problem of determining the projective representations among our representations
and suggest a candidate for a modular invariant bulk spectrum.

\section{Representations and their Structure}
\label{sec:repcont}

We begin with a quick review of minimal models and their generalisation to
logarithmic theories, for details on our notation please consult appendix
\ref{sec:notation}.
The Virasoro (non-logarithmic) minimal models and \(\mathcal{W}_{p,q}\)-models are labelled by coprime positive integers \((p,q)\)
and have central charge
\begin{align}
  c_{p,q}=1-6\frac{(p-q)^2}{pq}.
\end{align}
The irreducible representations have weights
\begin{align}\label{eq:infinitekac}
  h_{m,n}=\frac{(pn-qm)^2-(p-q)^2}{4pq},
\end{align}
where \(m\) and \(n\) are positive integers, and \(h_{m,n}=h_{p-m,q-n}\).

The non-logarithmic minimal models are representations of the vertex operator algebra (VOA) also known as the vacuum 
representation \(\mathcal{V}(h_{1,1}=0)\). This is the irreducible highest weight representation of the Virasoro algebra
based on the highest weight state \(\Omega\) with weight \(h=0\). The corresponding Verma module has two nullvectors
\(\mathcal{N}_1\) and \(\mathcal{N}_{(p-1)(q-1)}\) at levels 1 and \((p-1)(q-1)\) respectively. Setting both nullvectors
to zero one obtains the irreducible vacuum representation based on \(\Omega\). The highest weight representations
\(\mathcal{V}(h_{a,b})\) of the VOA with weight \(h_{a,b}\) are the representations of the Virasoro algebra for which the modes of the vertex
operators \(V(\mathcal{N}_1,z)\) and \(V(\mathcal{N}_{(p-1)(q-1)},z)\) act trivially.

The logarithmic theories of interest in this paper are constructed by only quotienting out the nullvector at level 1
in the Verma module corresponding to the VOA but not the nullvector at level \((p-1)(q-1)\). This prevents the the VOA
from being irreducible, but it is still indecomposable.
The corresponding theory is not rational, however, since the fusion of irreducible representations
of this VOA no longer closes on a finite set.
The repeated fusion of irreducibles produces an infinite series of irreducible representations with
weights of the form \eqref{eq:infinitekac} as well as reducible but indecomposable combinations of these 
irreducible representations.
To restore rationality the chiral algebra is
enlarged by three fields of conformal weight \((2p-1)(2q-1)\). We denote the resulting VOA by \(\mathcal{W}(p,q)\), its
irreducible highest weight representations of weight \(h\) are denoted 
by \(\mathcal{W}(h)\). 
The fusion of irreducible representations of \(\mathcal{W}(p,q)\) closes on a finite set, but apart from an
irreducible representation for every weight of the form \eqref{eq:weight} this
set also includes reducible but indecomposable combinations of these representations.

\subsection{Representation Content}
The \(\mathcal{W}_{p,q}\)-models close under the conjectured fusion rules of a grand total of 
\(4pq+13\frac{(p-1)(q-1)}{2}-2\) representations.
This is the smallest such set of representations, containing the vertex operator algebra (VOA) and the
irreducible representations.
For convenience we group these representations into two lists B and N. The
labelling of the weights in the following lists is explained in \eqref{eq:weight}.

\subsubsection*{Representations of Type B}
\begin{itemize}
\item \(2(p+q-1)\) irreducible representations:
  \begin{align}
    \mathcal{W}(h_{(r,q,\pm)})\qquad\textrm{and}\qquad\mathcal{W}(h_{(p,s,\pm)})
  \end{align}
  It was shown in \cite{Feigin:2006iv} that these irreducible representations can be interpreted as
  infinite sums of Virasoro representations.
\item \(4pq-2(p+q)\) rank 2 representations which are reducible but indecomposable and whose \(L_0\) action
  is not diagonalisable but rather contains \(2\times2\) Jordan blocks:
   \begin{align}\label{eq:r2}
     &\mathcal{R}^{(2)}(h_{(a,b,+)};h_{(p-a,b,-)})&
     &\mathcal{R}^{(2)}(h_{(a,b,+)};h_{(a,q-b,-)})\\
     &\mathcal{R}^{(2)}(h_{(a,q-b,-)};h_{(a,b,+)})&
     &\mathcal{R}^{(2)}(h_{(p-a,b,-)};h_{(a,b,+)})\nonumber\\
     &\mathcal{R}^{(2)}(h_{(a,q,+)})&
     &\mathcal{R}^{(2)}(h_{(p-a,b,-)})\nonumber\\
     &\mathcal{R}^{(2)}(h_{(p,b,+)})&
     &\mathcal{R}^{(2)}(h_{(p,q-b,-)})\nonumber
  \end{align}
  The first entry in \(\mathcal{R}^{(2)}(h_1;h_2)\) or \(\mathcal{R}^{(2)}(h)\) is the weight
  of the cyclic vector that generates the entire representation. For weights of type \(h_{(a,b,\pm)}\) this does
  not uniquely determine the rank 2 representation and an extra weight \(h_{(a',b',\mp)}\) is required to specify the
  representation in question.
\item \(2(p-1)(q-1)\) rank 3 representations which are reducible but indecomposable and whose \(L_0\) action
  is not diagonalisable but rather contains \(3\times3\) Jordan blocks:
  \begin{align}\label{eq:r3}
    &{\cal R}^{(3)}(h_{(a,b,\pm)})
    \nonumber
  \end{align}
  Here the argument \(h\) of \(\mathcal{R}^{(3)}(h)\) is the weight of the generating cyclic state.
\end{itemize}
Note that the rank 2 and 3 representations are obtained by repeated products of the irreducible representations.

\subsubsection*{Representations of type N}
\begin{itemize}
\item \(\frac12(p-1)(q-1)\) irreducible highest weight representations coming from the non-logarithmic minimal model:
  \begin{align}
    \mathcal{W}(h_{(a,b,0)})=\mathcal{W}(h_{(p-a,q-b,0)}).
  \end{align}
  It was shown in \cite{Feigin:2006iv} that these irreducible representations
  are just the
  irreducible Virasoro representations of the same weight.
\item \((p-1)(q-1)\) rank 1 highest weight representations
  \begin{align}
    \mathcal{W}_{a,b},
  \end{align}
  which are reducible but indecomposable
  and whose \(L_0\) action is diagonalisable. These representations were also introduced in \cite{Rasmussen:2008ez}.
\item \((p-1)(q-1)\) conjugates of the rank 1 representations \(\mathcal{W}_{a,b}\) which we shall denote by
  \begin{align}
    \mathcal{W}_{a,b}^\ast.
  \end{align}
\item \(2(p-1)(q-1)\) irreducible highest weight representations with weights
  that are descendants of
  those appearing in the non-logarithmic minimal models:
  \begin{align}
    \mathcal{W}(h_{(a,b,\pm)}).
  \end{align}
  It was shown in \cite{Feigin:2006iv} that these irreducible representations can be interpreted as
  infinite sums of Virasoro representations.
\end{itemize}

By the same arguments as in \cite{GabRunWoo} we believe that the
representations of type B are those that define a consistent boundary theory and therefore also show up in
lattice considerations such as \cite{Rasmussen:2008xi}, where fusion rules for
representations of this type are presented.
For these representations the notion of duals and contragredients is the same
(in fact these representations are self
contragredient and therefore also self dual); see \cite{GabRunWoo} section 3
for further details on duals and contragredients. 
It is also reassuring to note that this generalisation matches the representations appearing in 
\cite{Rasmussen:2008xi} exactly. 
By contrast the representations of type N do not define a consistent boundary theory, and, with the exception of the
representations of type \(\mathcal{W}_{a,b}\), the fusion rules for these representations are not yet known.
The goal of this paper is to extend the fusion rules to include all representations of type N. 

The representations of type \(\mathcal{W}(h_{(a,b,0)})\) are a somewhat special class of representations in N and have to 
be considered separately in a number of cases in
our analysis. We will therefore restrict ourselves to \(\textrm{N}_\times\), the set of all representations of types \(\mathcal{W}_{a,b},\ \mathcal{W}_{a,b}^\ast\) or
\(\mathcal{W}(h_{(a,b,\pm)})\), whenever we need to temporarily exclude the representations of type \(\mathcal{W}(h_{(a,b,0)})\) from our considerations.
In order to be able to extend the fusion rules to \({\rm N}_\times\), we need
to understand the detailed structure of representations of types \(\mathcal{W}_{a,b}\) and \(\mathcal{W}_{a,b}^\ast\) a little better:
\begin{itemize}
\item The representations of type \(\mathcal{W}_{a,b}\) correspond to weight \(h_{(a,b,0)}\) Verma modules
  where only the nullvector at level \(a b\) is quotiented out, but not the nullvector at level \((p-a)(q-b)\).
  They are characterised by the short exact sequences
  \begin{align}\label{eq:r1}
    \xymatrix{ 0\ar[r]&\mathcal{W}(h_{(a,b,+)})\ar[r]&\mathcal{W}_{a,b}\ar[r]&\mathcal{W}(h_{(a,b,0)})\ar[r]&0
      },
  \end{align}
  {\it i.e.}\ \(\mathcal{W}(h_{(a,b,+)})\) is a subrepresentation of \(\mathcal{W}_{a,b}\), which implies
  \begin{align}\label{eq:r1quot}
    {\cal W}(h_{(a,b,0)})={\cal W}_{a,b}/{\cal W}(h_{(a,b,+)}).
  \end{align}
  In particular the fusion of \(\mathcal{W}_{a,b}\) is the same as that of \(\mathcal{W}(h_{(a,b,0)})\) with all representations whose fusion with
  \(\mathcal{W}(h_{(a,b,+)})\) vanishes. This class of representations includes the VOA \(\mathcal{W}(p,q)\equiv\mathcal{W}_{1,1}\). 
\item The representations of type \(\mathcal{W}_{a,b}^\ast\) are generated by cyclic vectors of weight \(h_{(a,b,+)}\). These cyclic vectors
  are not highest weight however. We conjecture in analogy to \cite{GabRunWoo}, that the positive modes of the chiral 
  algebra map the cyclic vector to a vector that generates \(\mathcal{W}(h_{(a,b,0)})\) as a subrepresentation and that
  representations of type \(\mathcal{W}_{a,b}^\ast\) are characterised by the short exact sequences
  \begin{align}\label{eq:dr1}
    \xymatrix{ 0\ar[r]&\mathcal{W}(h_{(a,b,0)})\ar[r]&\mathcal{W}_{a,b}^\ast\ar[r]&\mathcal{W}(h_{(a,b,+)})\ar[r]&0
      }.
  \end{align}
    This implies
\begin{align}\label{eq:dr1quot}
  {\cal W}(h_{(a,b,+)})={\cal W}_{a,b}^\ast/{\cal W}(h_{(a,b,0)}).
\end{align}
In particular the fusion of \(\mathcal{W}_{a,b}^\ast\) is the same as that of \(\mathcal{W}(h_{(a,b,+)})\) with all representations whose fusion with
  \(\mathcal{W}(h_{(a,b,0)})\) vanishes.
\end{itemize}

\section{Fusion Rules}
\label{sec:fusionrules}

As mentioned in the introduction the fusion rules for representations of type B as well as
representations of type \(\mathcal{W}_{a,b}\) have already been determined in
\cite{Rasmussen:2008xi,Rasmussen:2008ez}. 
For the representations of type \(\mathcal{W}(h_{(a,b,0)})\) the fusion rules
are given by the minimal model fusion rules which have already been known for quite
some time \cite{Belavin:1984}.

We will now extend the the fusion rules to include all representations of type N, by first considering products of
representations of type \({\rm N}_\times\) with representations of types B or \(\textrm{N}_\times\), before considering representations of type
\(\mathcal{W}(h_{(a,b,0)})\).
The general strategy is to rewrite all representations in N as the the fusion product of a representations of type
\(\mathcal{W}_{a,b}\) and \(\mathcal{W}(h_{(1,1,0)}),\ \mathcal{W}(h_{(1,1,+)}),\ 
\mathcal{W}(h_{(1,1,-)})\) or \(\mathcal{W}_{1,1}^\ast\). Using commutativity and
associativity of the fusion product together with conjectured
fusion rules for \(\mathcal{W}(h_{(1,1,0)}),\ \mathcal{W}(h_{(1,1,+)}),\ 
\mathcal{W}(h_{(1,1,-)})\) and \(\mathcal{W}_{1,1}^\ast\) we can then define fusion rules for all of N.

\subsection{Products Involving Representations of Type $\textrm{N}_\times$}

\subsubsection{The Conjectured Fusion Rules for  $\mathcal{W}(h_{(1,1,+)}),\ \mathcal{W}(h_{(1,1,-)})$ and $\mathcal{W}_{1,1}^\ast$}

In order extend the fusion rules to representations of type \({\rm N}_\times\), we first need to understand the fusion rules of
\(\mathcal{W}(h_{(1,1,+)}),\ \mathcal{W}(h_{(1,1,-)})\) and \(\mathcal{W}_{1,1}^\ast\).
In analogy to \cite{GabRunWoo} and inspired by \cite{Feigin:2006iv} we conjecture 
that \(\mathcal{W}(h_{(1,1,+)}),\ 
\mathcal{W}(h_{(1,1,-)})\) and \(\mathcal{W}_{1,1}^\ast\) obey the following fusion rules
  \begin{align}\label{eq:compositionassumption}
    \mathcal{W}(h_{(1,1,+)})\otimes\mathcal{W}(h_{(1,1,+)})&=\mathcal{W}_{1,1}^\ast&
    \mathcal{W}(h_{(1,1,-)})\otimes\mathcal{W}(h_{(1,1,-)})&=\mathcal{W}_{1,1}^\ast\\
    \mathcal{W}(h_{(1,1,+)})\otimes\mathcal{W}(h_{(1,1,-)})&=\mathcal{W}(h_{(1,1,-)}).&
    \nonumber
  \end{align}
In appendix \ref{sec:consistency} we check \(\mathcal{W}(h_{(1,1,+)})\otimes\mathcal{W}(h_{(1,1,+)})\) for a number of cases using the NGK-algorithm introduced
in \cite{Nahm:1994by,Gaberdiel:1996kx}.
From \eqref{eq:compositionassumption} we can derive the remaining three products using associativity and the quotient
\eqref{eq:dr1quot}.
\begin{align}\label{eq:derivedcomposition}
  \mathcal{W}(h_{(1,1,+)})\otimes\mathcal{W}_{1,1}^\ast&\stackrel{\eqref{eq:dr1quot}}{=}
  \mathcal{W}(h_{(1,1,+)})\otimes\mathcal{W}(h_{(1,1,+)})= \mathcal{W}_{1,1}^\ast\\
  \mathcal{W}_{1,1}^\ast\otimes\mathcal{W}_{1,1}^\ast&\stackrel{\eqref{eq:compositionassumption}}{=}
  \mathcal{W}(h_{(1,1,+)})\otimes\mathcal{W}(h_{(1,1,+)})\otimes\mathcal{W}_{1,1}^\ast\\\nonumber
  &\stackrel{\eqref{eq:derivedcomposition}}{=}\mathcal{W}(h_{(1,1,+)})\otimes\mathcal{W}_{1,1}^\ast\stackrel{\eqref{eq:derivedcomposition}}{=}\mathcal{W}_{1,1}^\ast\\
  \label{eq:derivedcomposition2}
  \mathcal{W}_{1,1}^\ast\otimes\mathcal{W}(h_{(1,1,-)})&\stackrel{\eqref{eq:compositionassumption}}{=}
  \mathcal{W}(h_{(1,1,+)})\otimes\mathcal{W}(h_{(1,1,+)})\otimes\mathcal{W}(h_{(1,1,-)})\\
  &\stackrel{\eqref{eq:compositionassumption}}{=}\mathcal{W}(h_{(1,1,+)})\otimes\mathcal{W}(h_{(1,1,-)})
  =\mathcal{W}(h_{(1,1,-)})\nonumber
\end{align}
The use of the quotient \eqref{eq:dr1quot} is justified, because
\(\mathcal{W}(h_{(1,1,0)})\otimes\mathcal{W}(h_{(1,1,+)})=0\) as we will see in \eqref{eq:zerotimesrest}.
Again in analogy to \cite{GabRunWoo}, we conjecture that for representations of type \(\mathcal{W}_{a,b}\) the fusion with
the representations \(\mathcal{W}(h_{(1,1,+)}),\ 
\mathcal{W}(h_{(1,1,-)})\) and \(\mathcal{W}_{1,1}^\ast\) is given by
  \begin{align}\label{eq:factoringassumption}
    \mathcal{W}_{1,1}^\ast\otimes\mathcal{W}_{a,b}&=\mathcal{W}_{a,b}^\ast&
    \mathcal{W}(h_{(1,1,+)})\otimes\mathcal{W}_{a,b}&=\mathcal{W}(h_{(a,b,+)})\\\nonumber
    \mathcal{W}(h_{(1,1,-)})\otimes\mathcal{W}_{a,b}&=\mathcal{W}(h_{(a,b,-)}).&
  \end{align}
This is enough information to determine the product of the representations of
type \({\rm N}_\times\) using the fusion rules for two representations of type
\({\cal W}_{a,b}\) listed in appendix \ref{sec:Bfusion}.

To determine the product of a representation of type \({\rm N}_\times\) and a representation of type B,
it is sufficient to conjecture (also in analogy to \cite{GabRunWoo})
the action of \(\mathcal{W}(h_{(1,1,+)})\) and \(\mathcal{W}(h_{(1,1,-)})\) on the irreducible representations
  \begin{align}\label{eq:productassumption}
    \mathcal{W}(h_{(1,1,+)})\otimes\mathcal{W}(h_{(r,q,\pm)})&=\mathcal{W}(h_{(r,q,\pm)})&
    \mathcal{W}(h_{(1,1,+)})\otimes\mathcal{W}(h_{(p,s,\pm)})&=\mathcal{W}(h_{(p,s,\pm)})\\\nonumber
    \mathcal{W}(h_{(1,1,-)})\otimes\mathcal{W}(h_{(r,q,+)})&=\mathcal{W}(h_{(r,q,-)})&
    \mathcal{W}(h_{(1,1,-)})\otimes\mathcal{W}(h_{(p,s,+)})&=\mathcal{W}(h_{(p,s,-)}),
  \end{align}
since the rank 2 and rank 3 representations are products of the irreducible representations. On representations of type
B the action of
\(\mathcal{W}_{1,1}^\ast\) is the same as that of \(\mathcal{W}(h_{(1,1,+)})\), because of \eqref{eq:compositionassumption} and associativity.

In summary the \(3pq\) conjectured products \eqref{eq:compositionassumption},
\eqref{eq:factoringassumption} and \eqref{eq:productassumption} are sufficient to extend the fusion rules to 
\({\rm N}_\times\) by associativity and commutativity.

\subsubsection{Closed Fusion Formula for ${\rm N}_\times$}

In an attempt to improve readability we now introduce some more notation that will help us apply 
\eqref{eq:compositionassumption}, \eqref{eq:factoringassumption} and \eqref{eq:productassumption} to arbitrary products.
For every label \((a,b)\) we associate 4 representations of type \({\rm N}_\times\).
\begin{align}\label{eq:partners}
  \xymatrix{&\mathcal{W}_{a,b}\ar@<.2ex>@{-^{>}}[dr]^{\mathfrak{N}_-}\ar@<-.2ex>@{-_{>}}[dl]_{\mathfrak{N_+}}
    \ar@<.2ex>@{-^{>}}[dd]^(.75){\mathfrak{N}_\ast}\\
    \mathcal{W}(h_{(a,b,+)})\ar[rr]^(.65){\mathfrak{N}_-}\ar[dr]^{\mathfrak{N}_\ast}_{\mathfrak{N}_+}
    \ar@<-.2ex>@{-_{>}}[ur]_{\mathfrak{B}}&&
    \mathcal{W}(h_{(a,b,-)})\ar@<.2ex>@{-^{>}}[ul]^{\mathfrak{B}}\ar@<-.2ex>@{-_{>}}[dl]_{\mathfrak{N}_-}\\
    &\mathcal{W}_{a,b}^\ast
    \ar@<-.2ex>@{-_{>}}[ur]_{\mathfrak{N}_-}\ar@<.2ex>@{-^{>}}[uu]^(.75){\mathfrak{B}}
    }
\end{align}
The four labels \(\mathfrak{N}_{\pm,\ast}\) and \( \mathfrak{B}\) of the arrows are maps from the set 
of representations to itself, that are
linear with respect to direct sums, {\it i.e.}\ for a sum of representations they are evaluated for each representation
separately.
\begin{enumerate}
\item \(\mathcal{B}\) maps representations of type \({\rm N}_\times\) to
  their corresponding representation in \eqref{eq:partners} of type \(\mathcal{W}_{a,b}\)
  \begin{align}
    \mathfrak{B}(\mathcal{W}_{a,b})=\mathfrak{B}(\mathcal{W}_{a,b}^\ast)
    =\mathfrak{B}(\mathcal{W}(h_{(a,b,+)}))=\mathfrak{B}(\mathcal{W}(h_{(a,b,-)}))&=\mathcal{W}_{a,b},
  \end{align}
  and acts as identity for representations of type B.
  The image of \(\mathfrak{B}\) is therefore the representations of type B and representations
  of type \(\mathcal{W}_{a,b}\). For these representations the fusion rules have already been determined in
  \cite{Rasmussen:2008xi,Rasmussen:2008ez}.
\item The map \(\mathfrak{N}_\ast\) corresponds to the action of \(\mathcal{W}_{1,1}^\ast\) in \cite{GabRunWoo}, generalised for
  arbitrary \((p,q)\).
  It maps the representations in \eqref{eq:partners} to
  \begin{align}
    \mathfrak{N}_\ast(\mathcal{W}_{a,b})=\mathfrak{N}_\ast(\mathcal{W}_{a,b}^\ast)
    =\mathfrak{N}_\ast(\mathcal{W}(h_{(a,b,+)}))&=\mathcal{W}_{a,b}^\ast\\\nonumber
    \mathfrak{N}_\ast(\mathcal{W}(h_{(a,b,-)}))&=\mathcal{W}(h_{(a,b,-)}),
  \end{align}
   and acts as the identity on representations of type B. This corresponds
   precisely to the fusion products in in
  \eqref{eq:derivedcomposition}-\eqref{eq:derivedcomposition2}.
\item The map \(\mathfrak{N}_+\) corresponds to the action of \(\mathcal{W}(h_{(1,1,+)})\) 
  in \cite{GabRunWoo}, generalised for arbitrary \((p,q)\).
  It maps the representations in \eqref{eq:partners} to
  \begin{align}
    \mathfrak{N}_+(\mathcal{W}_{a,b})&=\mathcal{W}(h_{(a,b,+)})\\\nonumber
    \mathfrak{N}_+(\mathcal{W}_{a,b}^\ast)=\mathfrak{N}_+(\mathcal{W}(h_{(a,b,+)}))&=\mathcal{W}_{a,b}^\ast\\\nonumber
    \mathfrak{N}_+(\mathcal{W}(h_{(a,b,-)}))&=\mathcal{W}(h_{(a,b,-)}),
  \end{align}
  and acts as the identity on representations of type B. This corresponds
  precisely to the fusion products in \eqref{eq:compositionassumption}. 
\item The map \(\mathfrak{N}_-\) corresponds to the action of \(\mathcal{W}(h_{(1,1,1-)})\) in 
  \cite{GabRunWoo}, generalised for arbitrary \((p,q)\).
  It maps the representations in \eqref{eq:partners} to
  \begin{align}
    \mathfrak{N}_-(\mathcal{W}_{a,b})=\mathfrak{N}_-(\mathcal{W}_{a,b}^\ast)
    =\mathfrak{N}_-(\mathcal{W}(h_{(a,b,+)}))&=\mathcal{W}(h_{(a,b,-)})\\\nonumber
    \mathfrak{N}_-(\mathcal{W}(h_{(a,b,-)}))&=\mathcal{W}_{a,b}^\ast,
  \end{align}
    and on representations of type B it exchanges the weights \(h_{(r,s,\pm)}\) by \(h_{(r,s,\mp)}\)
  \begin{align*}
    \mathfrak{N}_-(\mathcal{W}(h_{(r,q,\pm)}))&=\mathcal{W}(h_{(r,q,\mp)})
    &\mathfrak{N}_-(\mathcal{W}(h_{(p,s,\pm)}))&=\mathcal{W}(h_{(p,s,\mp)})\\
    \mathfrak{N}_-(\mathcal{R}^{(2)}(h_{(r,q,\pm)}))&=\mathcal{R}^{(2)}(h_{(r,q,\mp)})
    &\mathfrak{N}_-(\mathcal{R}^{(2)}(h_{(p,s,\pm)}))&=\mathcal{R}^{(2)}(h_{(p,s,\mp)})\\
    \mathfrak{N}_-(\mathcal{R}^{(2)}(h_{(a,b,\pm)};h_{(a',b',\mp)}))&=\mathcal{R}^{(2)}(h_{(a,b,\mp)};h_{(a',b',\pm)})
    &\mathfrak{N}_-(\mathcal{R}^{(3)}(h_{(a,b,\pm)}))&=\mathcal{R}^{(3)}(h_{(a,b,\mp)}).
  \end{align*}
This corresponds
  precisely to the fusion products in \eqref{eq:compositionassumption}.
\end{enumerate}
By straight forward computation, we see that \(\mathfrak{N}_{\pm,\ast}\) and \(\mathfrak{B}\) satisfy the following
composition rules
\begin{align}
  \mathfrak{N}_\ast\circ\mathfrak{N}_\ast&=\mathfrak{N}_\ast&\mathfrak{N}_+\circ\mathfrak{N}_+&=\mathfrak{N}_\ast&
  \mathfrak{N}_-\circ\mathfrak{N}_-&=\mathfrak{N}_\ast
\end{align}
\begin{align*}
  \mathfrak{N}_\ast\circ\mathfrak{N}_+=\mathfrak{N}_+\circ\mathfrak{N}_\ast&=\mathfrak{N}_\ast&
  \mathfrak{N}_\ast\circ\mathfrak{N}_-=\mathfrak{N}_-\circ\mathfrak{N}_\ast&=\mathfrak{N}_-\\
  \mathfrak{N}_-\circ\mathfrak{N}_+=\mathfrak{N}_+\circ\mathfrak{N}_-&=\mathfrak{N}_-
\end{align*}
\begin{align*}
  \mathfrak{B}\circ\mathfrak{N}_{\pm,\ast}&=\mathfrak{B}&\mathfrak{N}_{\pm,\ast}\circ\mathfrak{B}=\mathfrak{N}_{\pm,\ast}.
\end{align*}
As a final piece of notation, when we write \(\mathfrak{N}_A\) for some representation \(A\), we mean
\begin{align}
  \mathfrak{N}_A=\left\{
    \begin{array}{cl}
      \mathfrak{N}_\ast&A=\mathcal{W}_{a,b}^\ast\\
      \mathfrak{N}_+&A=\mathcal{W}(h_{(a,b,+)})\\
      \mathfrak{N}_-&A=\mathcal{W}(h_{(a,b,-)})\\
      {\rm id}&\textrm{else}
    \end{array}\right..
\end{align}
This notation allows us to write
\begin{align}\label{eq:Nfactor}
  A=\mathfrak{N}_A\circ{\mathfrak{B}}(A).
\end{align}
This is just another way of writing
\eqref{eq:factoringassumption} in terms of \(\mathfrak{N}\) and \(\mathfrak{B}\).

Armed with all the information and notation of this section, we can then see that
the product of two indecomposable representations \(A\) and \(B\) (remember that we are not yet considering representations
of type \(\mathcal{W}(h_{(a,b,0)})\)) is given by
\begin{align}\label{eq:fusionrules}
  A\otimes B=
  \mathfrak{N}_A\circ\mathfrak{N}_B(\mathfrak{B}(A)\otimes\mathfrak{B}(B)).
\end{align}
In effect, this is simply rewriting the product of \(A\) and \(B\) in terms products involving
\(\mathcal{W}(h_{(1,1,+)}),\ \mathcal{W}(h_{(1,1,-)})\) and \(\mathcal{W}_{1,1}^\ast\) 
as well as the known product \(\mathfrak{B}(A)\otimes\mathfrak{B}(B)\). 
The product \(\mathfrak{B}(A)\otimes\mathfrak{B}(B)\)
 is evaluated using the fusion rules in 
\cite{Rasmussen:2008xi,Rasmussen:2008ez} and \(\mathfrak{N}_A\circ\mathfrak{N}_B\) is then applied to the
result.
As an example we will compute \(\mathcal{W}(h_{(1,2,+)})\otimes\mathcal{W}(h_{(1,2,-)})\) for \(q\geq3\). Using 
\eqref{eq:fusionrules} we find 
\begin{align*}
  \mathcal{W}(h_{(1,2,+)})\otimes\mathcal{W}(h_{(1,2,-)})&=\mathfrak{N}_{\mathcal{W}(h_{(1,2,+)})}
  \circ\mathfrak{N}_{\mathcal{W}(h_{(1,2,-)})}(\mathcal{W}_{1,2}\otimes\mathcal{W}_{1,2})=\mathfrak{N}_+\circ\mathfrak{N}_-
  (\mathcal{W}_{1,2}\otimes\mathcal{W}_{1,2})\\
  &=\mathfrak{N}_+\circ\mathfrak{N}_-(\mathcal{W}_{1,1}\oplus\mathcal{W}_{1,3})
  =\mathfrak{N}_-(\mathcal{W}_{1,1}\oplus\mathcal{W}_{1,3})\\
  &=\mathfrak{N}_-(\mathcal{W}_{1,1})\oplus\mathfrak{N}_-(\mathcal{W}_{1,3})
  =\mathcal{W}(h_{(1,1,-)})\oplus\mathcal{W}(h_{(1,3,-)}),
\end{align*}
where \(\mathcal{W}_{1,2}\otimes\mathcal{W}_{1,2}\) was evaluated using \eqref{eq:Bfusion}.
It is easy to check that \eqref{eq:fusionrules} agrees with 
\eqref{eq:compositionassumption}-\eqref{eq:productassumption}. We will show in section \ref{sec:associativity} that
it leads to associative fusion rules.

\subsection{Products Involving Representations of Type $\mathcal{W}(h_{(a,b,0)})$}
\label{sec:ab0}

The final step towards extending the fusion rules to all representations of type N, is to consider products involving
representations of type \(\mathcal{W}(h_{(a,b,0)})\).
As a first step we consider products of the form \(\mathcal{W}(h_{(1,1,0)})\otimes\mathcal{W}(h_{(r,s,\pm)})\). As mentioned before, the representations
of type \(\mathcal{W}(h_{(a,b,0)})\) come from the non-logarithmic minimal model and satisfy the minimal model fusion
rules among themselves
\begin{align}
  \mathcal{W}(h_{(a,b,0)})\otimes\mathcal{W}(h_{(a',b',0.)})=
  \sum_{\substack{k=1+|a-a'|\\k+a+a'=1\ {\rm mod}\ 2}}^{{\rm min}\{a+a'-1,2p-1-a-a'\}}\ 
  \sum_{\substack{l=1+|b-b'|\\l+b+b'=1\ {\rm mod}\ 2}}^{{\rm min}\{b+b'-1,2q-1-b-b'\}}
  \mathcal{W}(h_{(k,l,0)}).
\end{align}
Since \(\mathcal{W}(h_{(1,1,0)})\) is the VOA of the
non-logarithmic minimal model its fusion acts as the identity on representations in the non-logarithmic minimal model
and the product with any other irreducible representations vanishes. Therefore we have
\begin{align}\label{eq:zerotimesrest}
  \mathcal{W}(h_{(1,1,0)})\otimes\mathcal{W}(h_{(r,s,\pm)})=0.
\end{align}
This also ties in with what one would expect from the NGK-algorithm in \cite{Nahm:1994by,Gaberdiel:1996kx}
on the level of the
Virasoro algebra.\footnote{One sees that the quotient space of vectors not lying in the image of words with negative $L_0$-grading 
is at most 1-dimensional because
\(\mathcal{W}(h_{(1,1,0)})\) has the nullvector \({\cal N}_1=L_{-1}\Omega\) at level 1. Since the representations of type
\(\mathcal{W}(h_{(r,s,\pm)})\) are not representations of \(\mathcal{W}(h_{(1,1,0)})\) the remaining nullvectors will
impose additional constraints and the level 0 quotient must be zero.}

Using associativity and the fact that \(\mathcal{W}(h_{(1,1,0)})\) acts as the identity on representations of type
\(\mathcal{W}(h_{(a,b,0)})\) we can compute a more general version of \eqref{eq:zerotimesrest}
\begin{align}\label{eq:MMannihilation}
  \mathcal{W}(h_{(a,b,0)})\otimes\mathcal{W}(h_{(r,s,\pm)})
  =\mathcal{W}(h_{(a,b,0)})\otimes\mathcal{W}(h_{(1,1,0)})\otimes\mathcal{W}(h_{(r,s,\pm)})=0.
\end{align}
Since the rank 2 and 3 representations are just products of irreducible representations of type B, fusion of minimal
model representations \(\mathcal{W}(h_{(a,b,0)})\) with these vanishes as well. Therefore associativity guaranties that
\begin{align}
  \mathcal{W}(h_{(a,b,0)})\otimes\textrm{Representation of type B}=0.
\end{align}
This specifies the fusion rules of representations of type \(\mathcal{W}(h_{(a,b,0)})\) with all representations in B,
as well as the irreducible representations in N. All that remains is to describe
products of \(\mathcal{W}(h_{(a,b,0)})\) with \(\mathcal{W}_{a,b}\) or \(\mathcal{W}_{a.b}^\ast\).
Using associativity, \eqref{eq:derivedcomposition} and \eqref{eq:factoringassumption}
we see that \(\mathcal{W}(h_{(a,b,0)})\otimes\mathcal{W}_{a',b'}^\ast\) can be written as
\begin{align}
  \mathcal{W}(h_{(a,b,0)})\otimes\mathcal{W}_{a',b'}^\ast &=
  \mathcal{W}(h_{(a,b,0)})\otimes\mathcal{W}_{1,1}^\ast\otimes\mathcal{W}_{a',b'}\\\nonumber
  &=\mathcal{W}(h_{(a,b,0)})\otimes\mathcal{W}(h_{(1,1,+)})\otimes\mathcal{W}(h_{(1,1,+)})\otimes\mathcal{W}_{a',b'}=0
\end{align}
and therefore the fusion of all \(\mathcal{W}_{a',b'}^\ast\) with all \(\mathcal{W}(h_{(a,b,0)})\) vanishes.
Products of \(\mathcal{W}(h_{(a,b,0)})\) with representations of type \(\mathcal{W}_{a,b}\), can be computed
using the quotient \eqref{eq:r1quot}
\begin{align}\label{eq:MMrules}
  {\cal W}(h_{(a,b,0)})\otimes{\cal W}_{a',b'}={\cal W}(h_{(a,b,0)})\otimes{\cal W}(h_{(a',b',0)}).
\end{align}
In summary we therefore have that the fusion rules of \(\mathcal{W}(h_{(a,b,0)})\) satisfy:
\begin{enumerate}
\item All products of \({\cal W}(h_{(a,b,0)})\) with
  representations not of type \({\cal W}_{a,b}\) or \({\cal
    W}(h_{(a,b,0)})\) vanish.
\item Products of 
    \({\cal W}(h_{(a,b,0)})\) with representations of type \({\cal W}_{a,b}\) or \({\cal W}(h_{(a,b,0)})\) 
    are given by the non-logarithmic minimal model fusion rules.
\end{enumerate}

\subsection{Associativity}
\label{sec:associativity}

Now that we have extended the fusion rules to all representations of type N, we still need to prove that they are
associative. We do this by considering three cases. First we consider products of \(\mathcal{W}(h_{(a,b,0)})\)
with representations of type \({\cal W}_{a,b}\) or \({\cal W}(h_{(a,b,0)})\), secondly we consider products of
\(\mathcal{W}(h_{(a,b,0)})\) with anything else, and thirdly products not involving representations of type 
\(\mathcal{W}(h_{(a,b,0)})\).

\begin{enumerate}
\item As we discovered in the previous section, products involving a representation of type
    \({\cal W}(h_{(a,b,0)})\) together with representations of type \({\cal W}_{a,b}\) or \({\cal
      W}(h_{(a,b,0)})\)
    are given by the non-logarithmic minimal model fusion rules. These are known to be associative.
  \item As one can see from formula \eqref{eq:fusionrules} and the definitions of \(\mathfrak{N}_+,\ \mathfrak{N}_-\)
    and \(\mathfrak{N}_\ast\), products involving representations of type B, \(\mathcal{W}(h_{(a,b,\pm)})\) or 
      \(\mathcal{W}_{a,b}^\ast\) will never contain a representation of type \(\mathcal{W}_{a,b}\) or \(\mathcal{W}(h_{(a,b,0)})\) as a summand in 
      their result. Therefore all products involving representations of type \({\cal W}(h_{(a,b,0)})\)
  and representations not of type \({\cal W}_{a,b}\) or \({\cal
    W}(h_{(a,b,0)})\)
  vanish, regardless of the order in which the product is computed, hence this case is also 
  associative.
  \item Finally if we consider the product of three indecomposable representations \(A,B,C\) not of type
\(\mathcal{W}(h_{(a,b,0)})\), we then have
\begin{align}
  (A\otimes B)\otimes C
  &=(\mathfrak{N}_A\circ\mathfrak{N}_B(\mathfrak{B}(A)\otimes\mathfrak{B}(B)))\otimes C\\\nonumber
  &=\mathfrak{N}_A\circ\mathfrak{N}_B\circ\mathfrak{N}_C(\mathfrak{B}(\mathfrak{N}_A\circ\mathfrak{N}_B(\mathfrak{B}(A)
  \otimes\mathfrak{B}(B)))\otimes\mathfrak{B}(C))\\\nonumber
  &=\mathfrak{N}_A\circ\mathfrak{N}_B\circ\mathfrak{N}_C((\mathfrak{B}(A)\otimes\mathfrak{B}(B))\otimes\mathfrak{B}(C))\\
  \nonumber
  &=\mathfrak{N}_A\circ\mathfrak{N}_B\circ\mathfrak{N}_C(\mathfrak{B}(A)\otimes\mathfrak{B}(B)\otimes \mathfrak{B}(C)),\\
  A\otimes(B\otimes C)&=A\otimes(\mathfrak{N}_B\circ\mathfrak{N}_C(\mathfrak{B}(B)\otimes \mathfrak{B}(C)))\\\nonumber
  &=\mathfrak{N}_A\circ\mathfrak{N}_B\circ\mathfrak{N}_C(\mathfrak{B}(A)\otimes\mathfrak{B}(\mathfrak{N}_B\circ
  \mathfrak{N}_C(\mathfrak{B}(B)\otimes \mathfrak{B}(C))))\\\nonumber
  &=\mathfrak{N}_A\circ\mathfrak{N}_B\circ\mathfrak{N}_C(\mathfrak{B}(A)\otimes(\mathfrak{B}(B)\otimes \mathfrak{B}(C)))\\
  \nonumber
  &=\mathfrak{N}_A\circ\mathfrak{N}_B\circ\mathfrak{N}_C(\mathfrak{B}(A)\otimes\mathfrak{B}(B)\otimes \mathfrak{B}(C)).
  \nonumber
\end{align}
\end{enumerate}
Therefore the fusion rules defined in this section are associative if the fusion rules in 
\cite{Rasmussen:2008xi,Rasmussen:2008ez} are associative.\footnote{We have checked this
explicitly for $(p,q)=(2,3)$ in \cite{GabRunWoo} and this is believed to be true for all $(p,q)$.}

\section{The Grothendieck Group}
\label{sec:grothendieckgroup}

We will now study the Grothendieck group, an object closely related to open string spectra.
The Grothendieck group \(K_0\equiv K_0({\rm Rep}(\mathcal{W}(p,q)))\) of representations of \(\mathcal{W}(p,q)\) is,
roughly speaking, the quotient set obtained by identifying two representations if they have the same character.
We denote the equivalence class of a representation \(\mathcal{R}\) by \([\mathcal{R}]\). The group operation is abelian
and defined by the direct sum
\begin{align}
  [\mathcal{R}_1]+[\mathcal{R}_2]=[\mathcal{R}_2\oplus\mathcal{R}_2].
\end{align}
For example the exact sequences \eqref{eq:r1} and \eqref{eq:dr1} imply
\begin{align}
  [\mathcal{W}_{a,b}]=[\mathcal{W}_{a,b}^\ast]=[\mathcal{W}(h_{(a,b,0)})]+[\mathcal{W}(h_{(a,b,+)})].
\end{align}
Since the characters of all indecomposable representations can be written as linear combinations of characters of
irreducible representations (see appendix \ref{sec:chars}) the Grothendieck group is the free abelian group generated
by the irreducible representations.

For non-logarithmic rational conformal field theories, the Grothendieck group also has a product structure turning it
into a ring which is defined by
\begin{align}
  [{\cal R}_1]\cdot[{\cal R}_2]=[{\cal R}_1\otimes {\cal R}_2].
\end{align}
For the \(\mathcal{W}_{p,q}\) triplet models the situation is not quite as simple, a consistent product structure
can no longer be defined for the entire Grothendieck group. The counter
example in \cite{GabRunWoo} can be easily generalised for all \((p,q)\):
\begin{align}
  [\mathcal{W}(h_{(1,1,0)})]\cdot[\mathcal{W}_{a,b}^\ast]&=[\mathcal{W}(h_{(1,1,0)})\otimes\mathcal{W}_{a,b}^\ast]=0
  \quad{\rm versus}\\
  [\mathcal{W}(h_{(1,1,0)})]\cdot[\mathcal{W}_{a,b}^\ast]&=[\mathcal{W}(h_{(1,1,0)})]\cdot\left(
    [\mathcal{W}(h_{(a,b,0)})]+[\mathcal{W}(h_{(a,b,+)})]\right)\nonumber\\
  &=[\mathcal{W}(h_{(1,1,0)})\otimes\mathcal{W}(h_{(a,b,0)})]+[\mathcal{W}(h_{(1,1,0)})\otimes\mathcal{W}(h_{(a,b,+)})]
  =[\mathcal{W}(h_{(a,b,0)})].\nonumber
\end{align}

It was shown in \cite{GabRunWoo}, however that if a representation \({\cal M}\) has a dual it induces a well-defined map
\begin{align}
  K_0&\rightarrow K_0\\
  [{\cal R}]&\mapsto[{\cal M}\otimes {\cal R}].\nonumber
\end{align}
We can therefore define the subgroup \(K_0^r\) of \(K_0\) generated by \([{\cal R}]\) for all \({\cal R}\) which have
a dual representation. As in \cite{GabRunWoo} we believe that these are the representations of type B together with
the representations of type \(\mathcal{W}_{a,b}\).
\(K_0^r\) is then spanned by \(\frac12\left(5pq-(p+q)+1\right)\) classes of representations
\begin{align}
  K_0^r:={\rm span}_{\mathbb{Z}}([{\cal W}_{a,b}],[{\cal W}(h_{(r,q,\pm)})],[{\cal
    W}(h_{(q,s,\pm)})], [{\cal
    R}^{(2)}(h_{(a,b,+)};h_{(p-a,b,-)})],\\
  [{\cal R}^{(2)}(h_{(p-a,b,-)};h_{(a,b,+)})],[{\cal R}^{(2)}(h_{(a,q-b,-)};h_{(a,b,+)})]).\nonumber
\end{align}
This is less than the total number of representations of types B and \(\mathcal{W}_{a,b}\) since their characters are
linearly dependent as one can see in appendix \ref{sec:chars}.
The basis can also be written in terms of irreducible representations, but
the product then no longer corresponds to fusion. Rather one has to first perform the following substitutions
before interpreting the product as fusion
\begin{align}
  [{\cal W}(h_{(a,b,0)})]&=[{\cal
    R}^{(2)}(h_{(a,b,+)};h_{(p-a,b,-)})]-[{\cal
    R}^{(2)}(h_{(p-a,b,-)};h_{(a,b,+)})]\\\nonumber
  [{\cal W}(h_{(a,b,+)})]&=[{\cal W}_{a,b}]-[{\cal W}(h_{(a,b,0)})]\\\nonumber
  &=[{\cal W}_{a,b}]+[{\cal
    R}^{(2)}(h_{(p-a,b,-)};h_{(a,b,+)})]-[{\cal
    R}^{(2)}(h_{(a,b,+)};h_{(p-a,b,-)})]\\\nonumber
  2[{\cal W}(h_{(a,b,-)})]&=[{\cal
    R}^{(2)}(h_{(a,b,-)};h_{(p-a,b,+)})]-2[{\cal W}(h_{(p-a,b,+)})]\\\nonumber
  &=[{\cal R}^{(2)}(h_{(a,b,-)};h_{(p-a,b,+)})]+2[{\cal R}^{(2)}(h_{(p-a,b,+)};h_{(a,b,-)})]\\\nonumber
  &\quad\ -2[{\cal R}^{(2)}(h_{(a,b,-)};h_{(p-a,b,+)})]-2[{\cal W}_{p-a,b}].
\end{align}
For example the square of \([\mathcal{W}(h_{(1,1,0)})]\) is then given by
\begin{align}
  [\mathcal{W}(h_{(1,1,0)})]\cdot[\mathcal{W}(h_{(1,1,0)})]&=
  ([{\cal R}^{(2)}(h_{(1,1,+)};h_{(p-1,1,-)})]-[{\cal R}^{(2)}(h_{(p-1,1,-)};h_{(1,1,+)})])\\
  \nonumber
  &\quad\ \cdot([{\cal R}^{(2)}(h_{(1,1,+)};h_{(p-1,1,-)})]-[{\cal R}^{(2)}(h_{(p-1,1,-)};h_{(1,1,+)})])\\
  \nonumber
  &=[{\cal R}^{(2)}(h_{(1,1,+)};h_{(p-1,1,-)})^2]+[{\cal R}^{(2)}(h_{(p-1,1,-)};h_{(1,1,+)})^2]\\
  \nonumber
  &\quad\ -2[{\cal R}^{(2)}(h_{(1,1,+)};h_{(p-1,1,-)})\otimes{\cal R}^{(2)}(h_{(p-1,1,-)};h_{(1,1,+)})]\nonumber\\
  &=0,\nonumber
\end{align}
where the fusion products where evaluated using the rules in \cite{Rasmussen:2008xi}.

By a long but straightforward computation one sees that
\begin{align}
  [{\cal W}(h_{(a,b,0)})]\cdot[{\cal W}(h_{(r,s,\mu)})]=\left\{
    \begin{array}{l}
      \mu\cdot\sum^{p-|a+r-p|-1}_{i=|r-a|+1,{\rm by
          2}}\sum^{q-|q-b-s|-1}_{j=|b-s|+1,{\rm by 2}}[{\cal
        W}(h_{(i,j,0)})]\\
      \qquad{\rm if }\
      (r,s)\in\{1,\dots,p-1\}\times\{1,\dots,q-1\}\\
      0\quad{\rm otherwise}
    \end{array}
\right.,
\end{align}
{\it i.e.}\ the classes of representations of type \(\mathcal{W}(h_{(a,b,0)})\) form the ideal
\begin{align}
  {\cal I}_0=\bigoplus_{(a,b)\in J}\mathbb{Z}[\mathcal{W}(h_{(a,b,0)})],
\end{align}
where
\begin{align}
  J:=\{(a,b)|1\leq a\leq p-1,1\leq b\leq q-1,qa+ps\leq pq\}.
\end{align}
It is important to remember, that a number of rank 2 and 3 representations have the same characters and therefore
belong in the same equivalence class, while performing this computation.

The Grothendieck group has a direct interpretation in terms of cylinder diagrams. It is therefore interesting to
consider the subgroup \(K_0^b\) generated by representations corresponding to boundary
conditions. 
The open string
spectrum between two boundaries labelled by representations \(A\) and \(B\) is given by
\begin{align}
  Z(q)_{A\rightarrow B}={\rm tr}_{B\otimes A^\ast}(q^{L_0-c/24}),
\end{align}
the character of \(B\otimes A^\ast\), or more formally it only depends on the class \([B\otimes A^\ast]\).
Restricting ourselves to the equivalence classes of representations of type B, {\it i.e.}\ drop 
\([{\cal W}_{a,b}]\), \(K_0^b\) is spanned by \(2pq\) generators
\begin{align}
  K_0^b={\rm span}_{\mathbb{Z}}&([{\cal W}(h_{(r,q,\pm)})],[{\cal
    W}(h_{(p,s,\pm)})])\oplus\mathcal{I}_0\\\nonumber
  &\bigoplus_{(a,b)\in J}\left(2\mathbb{Z}([{\cal W}(h_{(a,b,+)})]+[{\cal
      W}(h_{(a,q-b,-)})])\right.\\\nonumber
  &\oplus 2\mathbb{Z}([{\cal W}(h_{(a,b,+)})]+[{\cal
      W}(h_{(p-a,b,-)})])\\\nonumber
  &\left.\oplus\, 2\mathbb{Z}([{\cal W}(h_{(p-a,q-b,+)})]+[{\cal W}(h_{(a,q-b,-)})])\right).
\end{align}
Since all representations of B have duals and close under fusion, \(K_0^b\)
also closes under the product induced by fusion.

\section{Projective Representations and Modular Invariant Partition Functions}
\label{sec:projective}

In this section we will look for projective representations. These are of
particular interest to us, since it is believed \cite{Gaberdiel:2007jv,Quella:2007hr} that the bulk spectrum
of these theories should be describable in terms of a quotient of
\begin{align}
  \bigoplus_i{\cal P}_i\otimes_{\mathbb{C}}\overline{\cal P}_i,
\end{align}
where the sum runs over all projective representations and the bar
refers to right-movers.

Before we determine which of our \(\mathcal{W}(p,q)\) representations are projective, we recall one
of a number of equivalent definitions of projective representations.
\begin{Def}
  A \(\mathcal{W}(p,q)\) representation \(\mathcal{P}\) is \emph{projective}, if
  given an intertwiner \(f:\mathcal{P}\rightarrow\mathcal{M}'\) and a surjective intertwiner
  \(g:\mathcal{M}\rightarrow\mathcal{M}'\), there exits an intertwiner \(e:\mathcal{P}\rightarrow\mathcal{M}\)
  making the following diagram commute.
  \begin{align}\label{eq:proj}
    \xymatrix{&\mathcal{P}\ar[d]_f\ar[dl]_e\\
    \mathcal{M}\ar[r]_g&\mathcal{M}'}
  \end{align}
\end{Def}

The irreducible representations \({\cal W}(h_{(p,q,\pm)})\) do not share weights with any other \(\mathcal{W}(p,q)\) representations
in our lists B and N, therefore they can only have non trivial intertwiners
with themselves. This makes them promising candidates for being projective.
So if we set \(\mathcal{P}=\mathcal{W}(h_{(p,q,\pm)})\) in diagram
\eqref{eq:proj} and \(\mathcal{M}'\neq{\cal W}(h_{(p,q,\pm)}))\) we have \(f\equiv0\) and diagram \eqref{eq:proj} commutes for \(e\equiv 0\). If on the other hand
\(\mathcal{P}=\mathcal{W}(h_{(p,q,\pm)})=\mathcal{M}'\) then by Schur's lemma \(f=c_f\cdot{\rm id},\ c_f\in\mathbb{C}\) and the only \(\mathcal{M}\) for which \(g\) can be surjective is \(\mathcal{W}(h_{(p,q,+)})\)
with \(g=c_g\cdot{\rm id},\ c_g\in\mathbb{C}\setminus\{0\}\). Therefore the
diagram \eqref{eq:proj} commutes for \(e=\frac{c_f}{c_g}\cdot{\rm id}\)
 and the 2 representations
\(\mathcal{W}(h_{(p,q,\pm)})\) are projective.

A further property of projective representations is that their products with representations that have duals
are also projective. We assume that in analogy to \cite{GabRunWoo} the representations of type B, \(\mathcal{W}_{a,b}\)
and \(\mathcal{W}_{a,b}^\ast\) have duals. By computing the product of all representations of type B, \(\mathcal{W}_{a,b}\)
and \(\mathcal{W}_{a,b}^\ast\) with \(\mathcal{W}(h_{(p,q,\pm)})\)
we find \(2pq\) indecomposable representations that also ought to be projective. We denote these representations by
\(\mathcal{P}(h)\) where \(h\) is the weight of the irreducible representation they are a cover of:
\begin{enumerate}
\item Irreducible representations
  \begin{align*}
    {\cal W}(h_{(p,q,\pm)})&={\cal P}(h_{(p,q,\pm)})
  \end{align*}
\item Rank 2 representations
  \begin{align*}
    {\cal R}^{(2)}(h_{(a,q,\pm)})&={\cal P}(h_{(a,q,\pm)})\\
    {\cal R}^{(2)}(h_{(p,b,\pm)})&={\cal P}(h_{(p,b,\pm)})
  \end{align*}
\item Rank 3 representations
  \begin{align*}
    {\cal R}^{(3)}(h_{(a,b,\pm)})&={\cal P}(h_{(a,b,\pm)})
  \end{align*}
\end{enumerate}
This accounts for the projective covers of all representations in B and N, except for \(\mathcal{W}(h_{(a,b,0)})\).
In fact, none of the representations in B and N appears to be a projective cover of \(\mathcal{W}(h_{(a,b,0)})\).

\subsection{Modular Invariant Partition Function}

The modular transformation properties of the characters as well as a modular
invariant combination of these characters are given in \cite{Feigin:2006iv}. It was discovered in
\cite{GabRunWoo} that for \((p,q)=(2,3)\) this modular invariant function can
be written as
\begin{align*}
  Z_{p,q}=&\sum_{h_{(r,s,\pm)}}\dim({\rm Hom}({\cal P}(h_{(r,s,\pm)}),{\cal
    P}(h_{(r,s,\pm)})))^{-2}\cdot\left|\chi[{\cal P}(h_{(r,s,\pm)})](q)\right|^2,
\end{align*}
where \({\rm Hom}(U,W)\) is the space of intertwiners from \(U\) to \(W\).
We conjecture following the arguments of \cite{GabRunWoo} that the relation
\begin{align}
  {\rm Hom}(U,V)\cong {\rm Hom}(U\otimes V^\ast,\mathcal{W}_{1,1}^\ast)
\end{align}
holds for the spaces of intertwiners between representations for general
\((p,q)\) and can therefore be used to calculate
\(\dim{\rm Hom}(\mathcal{P}(h_{(r,s,\pm)}),\mathcal{P}(h_{(r,s,\pm)}))\) using (also in analogy to \cite{GabRunWoo})
\begin{align}
  \dim{\rm Hom}(U,\mathcal{W}_{1,1}^\ast)=\left\{
    \begin{array}{cl}
      1&U\in\{\mathcal{W}(h_{(1,1,0)}),\mathcal{W}_{1,1},\mathcal{W}_{1,1}^\ast,\mathcal{W}_{p-1,q-1},
      \mathcal{R}^{(3)}(h_{(1,1,+)})\},\\
      &\qquad\ \mathcal{R}^{(2)}(h_{(1,1,+)};h_{(p-1,1,-)}),\mathcal{R}^{(2)}(h_{(1,1,+)};h_{(1,q-1,-)})\}\\
      0&{\rm else}
    \end{array}\right..
\end{align}
We find that
\begin{align}
  \dim{\rm Hom}(\mathcal{P}(h_{(p,q,\pm)}),\mathcal{P}(h_{(p,q,\pm)}))&=1&
  \dim{\rm Hom}(\mathcal{P}(h_{(r,q,\pm)}),\mathcal{P}(h_{(r,q,\pm)}))&=2\\
  \dim{\rm Hom}(\mathcal{P}(h_{(p,s,\pm)}),\mathcal{P}(h_{(p,s,\pm)}))&=2&
  \dim{\rm Hom}(\mathcal{P}(h_{(a,b,\pm)}),\mathcal{P}(h_{(a,b,\pm)}))&=4.\nonumber
\end{align}
These values for the dimension of the \({\rm Hom}\)-spaces are also consistent
with the conjectured embedding structures in \cite{Rasmussen:2008xi}.
We thus have
\begin{align}
  Z_{p,q}=&\left|\chi[{\cal P}(h_{(p,q,+)})](q)\right|^2+\left|\chi[{\cal
      P}(h_{(p,q,-)})](q)\right|^2\\\nonumber
  &+\frac14\left(\sum_{r=1}^p\left|\chi[{\cal P}(h_{(r,q,+)})](q)\right|^2+\left|\chi[{\cal
      P}(h_{(r,q,-)})](q)\right|^2\right)\\\nonumber
  &+\frac14\left(\sum_{s=1}^q\left|\chi[{\cal P}(h_{(p,s,+)})](q)\right|^2+\left|\chi[{\cal
      P}(h_{(p,s,-)})](q)\right|^2\right)\\\nonumber
  &+\frac{1}{16}\left(\sum_{\substack{1\leq a\leq p\\1\leq b\leq q}}\left|\chi[{\cal P}(h_{(a,b,+)})](q)\right|^2
    +\left|\chi[{\cal
      P}(h_{(a,b,-)})](q)\right|^2\right).
\end{align}
We have convinced ourselves of the modular invariance of the formula above, by extensive numerical
checks, but unfortunately the general expression seems to be to unwieldy for computer algebra systems to
handle and we have not yet found a way to verify it for all coprime pairs \((p,q)\).

\section{Conclusions}

In this paper we have studied the \(\mathcal{W}_{p,q}\) triplet models. The structure we have found is very analogous to the results
already obtained for the \({\cal W}_{2,3}\) models in \cite{GabRunWoo}.
The representations appearing in \cite{GabRunWoo} were generalised
for arbitrary \((p,q)\) and we showed that the fusion rules can easily be extended to these new representations
 by conjecturing very plausible fusion rules for
\(\mathcal{W}_{1,1}^\ast, \mathcal{W}(h_{(1,1,+)})\) and \(\mathcal{W}(h_{(1,1,-)})\) as well as using associativity and
commutativity of the fusion product.
Subsequently the Grothendieck group \(K_0\) was constructed together with subgroups \(K_0^r\) and \(K_0^b\) on which
consistent fusion induced products can be defined.
As a final exercise the projective representations where identified and used to suggest the structure of a modular
invariant bulk theory.

This entire representations theoretic analysis suggests that a consistent boundary theory can be defined from which
one can then construct a bulk theory in analogy to the Cardy case 
\cite{Runkel:1998pm,Felder:1999mq,tft1,unique,Kong:2008ci} or as was done for the \(\mathcal{W}_{1,p}\)
models in \cite{Gaberdiel:2007jv}. It would be very interesting to 
construct the bulk theory for example for \((p,q)=(2,3)\). If this succeeds this should probably
directly generalise to \((p,q)\) as the analysis of the fusion rules in this paper did.

\subsection*{Acknowledgements}
The author would like to thank Matthias Gaberdiel and Ingo Runkel for many helpful discussions
and comments as well as J\o rgen Rasmussen for helpful comments.
This research is supported by the Swiss National Science Foundation.

\subsection*{Note Added}

While this paper was being written another paper with significant overlap appeared on the arXiv \cite{Rasmussen:2009}.
In \cite{Rasmussen:2009} the same fusion algebra is computed from a different
perspective by focusing on symmetry principles. The general
philosophy in this paper however, was to use associativity and commutativity
to reexpress arbitrary fusion products as known products
and products involving \({\cal W}_{1,1}^\ast,\ {\cal W}(h_{(1,1,+)})\)
and \({\cal W}(h_{(1,1,-)})\) for which we conjectured fusion rules.

\appendix

\section{Notation}
\label{sec:notation}

Unless stated otherwise we will always assume that
\begin{align*}
  \alpha&\in\{0,\dots,p\},&r&\in\{1,\dots,p\}&a&\in\{1,\dots,p-1\}\\
  \beta&\in\{0,\dots,q\},&s&\in\{1,\dots,q\}&b&\in\{1,\dots,q-1\}
\end{align*}
 We will be using the notation of
\cite{Feigin:2006iv} to compactly label the weights
\begin{align}\label{eq:weight}
  h_{(a,b,0)}&:=h_{a,b}=h_{p-a,q-b}\\
  h_{(r,s,+)}&:=h_{2p-r,s}=h_{r,2q-s}=h_{r,s}+(p-r)(q-s)\nonumber\\
  h_{(r,s,-)}&:=h_{3p-r,s}=h_{r,3q-s}=h_{r,s}+(p-r)(q-s)+\frac54pq-\frac{ps+qr}{2}\nonumber.
\end{align}

\section{Dictionary to the Notation in other Works}

\subsection*{The Notation in \cite{Rasmussen:2008xi,Rasmussen:2008ez}}

The fusion rules for the image of \(\mathfrak{B}\) and representations of type \(\mathcal{W}_{a,b}\)
are contained in \cite{Rasmussen:2008xi,Rasmussen:2008ez}.
Here we give a dictionary between the two notations

\begin{center}
\begin{tabular}{c|c}
  our notation & notation in \cite{Rasmussen:2008xi,Rasmussen:2008ez} \\
  \hline
  \({\cal W}_{a,b}\)&\((a,b)\)\\
  \({\cal W}(h_{(r,q,-(-)^\kappa)})\)&\((r,\kappa q)\)\\
  \({\cal W}(h_{(p,s,-(-)^\kappa)})\)&\((\kappa p,s)\)\\
  \({\cal R}^{(2)}(h_{(p-a,b,-(-)^\kappa)};h_{(a,b,(-)^\kappa)})\)&\({\cal R}^{a,0}_{\kappa p,b}\)\\
  \({\cal R}^{(2)}(h_{p-a,q,-(-)^\kappa})\)&\({\cal R}^{a,0}_{\kappa p,q}\)\\
  \({\cal R}^{(2)}(h_{(a,q-b,-(-)^\kappa)};h_{(a,b,(-)^\kappa)})\)&\({\cal R}^{0,b}_{a,\kappa q}\)\\
  \({\cal R}^{(2)}(h_{(p,q-b,-(-)^\kappa)})\)&\({\cal R}^{0,b}_{p,\kappa q}\)\\
  \({\cal R}^{(3)}(h_{(p-a,q-b,+)})\)&\({\cal R}^{a,b}_{p,q}\)\\
  \({\cal R}^{(3)}(h_{(p-a,q-b,-)})\)&\({\cal R}^{a,b}_{p,2q}\)
\end{tabular}
\end{center}

\noindent
Here \(\kappa=1,2\).
The representations of type $\mathcal{W}_{a,b}^\ast$, $\mathcal{W}(h_{(a,b,\pm)})$ and $\mathcal{W}(h_{(a,b,0)})$  
are not considered in \cite{Rasmussen:2008xi,Rasmussen:2008ez}.

\subsection*{The Notation in \cite{Feigin:2006iv}}

\begin{center}
\begin{tabular}{c|c}
  our notation & notation in \cite{Feigin:2006iv} \\
  \hline
  \(\mathcal{W}_{a,b}\)&\( \mathcal{K}_{a,b}^+ \) \\
  \(\mathcal{W}(h_{(a,b,0)})\)& \(\mathcal{X}_{a,b}\) \\
  \(\mathcal{W}(h_{(a,b,+)})\)& \(\mathcal{X}_{a,b}^+\)  \\
  \(\mathcal{W}(h_{(a,b,-)})\)& \(\mathcal{K}_{a,b}^- = \mathcal{X}_{a,b}^-\)\\
  \(\mathcal{W}(h_{(r,q,\pm)})\)& \(\mathcal{K}_{r,q}^\pm = \mathcal{X}_{r,q}^\pm\)\\
  \(\mathcal{W}(h_{(p,s,\pm)})\)&\(\mathcal{K}_{p,s}^\pm = \mathcal{X}_{p,s}^\pm\)
\end{tabular}
\end{center}

\noindent
The representations of type $\mathcal{W}_{a,b}^\ast$ as well as the rank 2 and 3 representations
are not considered in \cite{Feigin:2006iv}.

\subsection*{The Notation in \cite{GabRunWoo}}

\begin{center}
\begin{tabular}{c|c}
  our notation & notation in \cite{GabRunWoo} \\
  \hline
  \(\mathcal{W}_{1,1}\)&\(\mathcal{W}\)\\
  \(\mathcal{W}_{1,1}^\ast\)&\(\mathcal{W}^\ast\)\\
  \(\mathcal{W}_{1,2}\)&\(\mathcal{Q}\)\\
  \(\mathcal{W}_{1,2}^\ast\)&\(\mathcal{Q}^\ast\)\\
  \(\mathcal{R}^{(2)}(h_{(a,b,+)};h_{(p-a,b,-)})\)&\({\cal R}^{(2)}(h_{(a,b,0)},h_{(a,b,+)})_{h_{(p-a,b,-)}}\)\\
  \(\mathcal{R}^{(2)}(h_{(a,b,+)};h_{(a,q-b,-)})\)&\({\cal R}^{(2)}(h_{(a,b,0)},h_{(a,b,+)})_{h_{(a,q-b,-)}}\)\\
  \(\mathcal{R}^{(2)}(h_{(a,q-b,-)};h_{(a,b,+)})\)&\({\cal R}^{(2)}(h_{(a,b,+)},h_{(a,q-b,-)})\)\\
  \(\mathcal{R}^{(2)}(h_{(p-a,b,-)};h_{(a,b,+)})\)&\({\cal R}^{(2)}(h_{(a,b,+)},h_{(p-a,b,-)})\)\\
  \(\mathcal{R}^{(2)}(h_{(a,q,+)})\)&\({\cal R}^{(2)}(h_{(a,q,+)},h_{(a,q,+)})\)\\
  \(\mathcal{R}^{(2)}(h_{(p-a,b,-)})\)&\({\cal R}^{(2)}(h_{(a,q,+)},h_{(p-a,q,-)})\)\\
  \(\mathcal{R}^{(2)}(h_{(p,b,+)})\)&\({\cal R}^{(2)}(h_{(p,b,+)},h_{(p,b,+)})\)\\
  \(\mathcal{R}^{(2)}(h_{(p,q-b,-)})\)&\({\cal R}^{(2)}(h_{(p,b,+)},h_{(p,q-b,-)})\)\\
  \({\cal R}^{(3)}(h_{(a,b,+)})\)&\({\cal R}^{(3)}(h_{(a,b,0)},h_{(a,b,0)},h_{(a,b,+)},h_{(a,b,+)})\)\\
  \({\cal R}^{(3)}(h_{(a,q-b,-)})\)&\({\cal R}^{(3)}(h_{(a,b,0)},h_{(p-a,q-b,+)},h_{(a,b,+)},h_{(a,q-b,-)})\)
\end{tabular}
\end{center}

\section{Fusion for Representations of Type ${\cal W}_{a,b}$}
\label{sec:Bfusion}

In order to have some fusion rules at hand, we include the rules for products of representations of type
\(\mathcal{W}_{a,b}\) in our notation, please refer to \cite{Rasmussen:2008xi,Rasmussen:2008ez} for the remaining rules.
\begin{align}\label{eq:Bfusion}
  \mathcal{W}_{a,b}&\otimes\mathcal{W}_{a',b'}=
  \bigoplus^{p-|p-a-a'|-1}_{\substack{i=|a-a'|+1\\\textrm{ by }2}}\ 
  \bigoplus^{q-|q-b-b'|-1}_{\substack{j=|b-b'|+1\\\textrm{ by }2}}
  \mathcal{W}_{i,j}\\\nonumber
  &\oplus\bigoplus^{a+a'-p-1}_{\substack{\alpha=a+a'-p-1 \textrm{ mod 2}\\\textrm{by 2}}}\ 
  \bigoplus^{q-|q-b-b'|-1}_{\substack{j=|b-b'|+1\\\textrm{by }2}}
   \mathcal{R}^{(2)}(h_{(p-\alpha,j,+)};h_{(\alpha,j,-)})
  \\\nonumber
  &\oplus\bigoplus^{b+b'-q-1}_{\substack{\beta=b+b'-q-1 \textrm{ mod 2}\\ \textrm{by 2}}}\ 
  \bigoplus^{p-|p-a-a'|-1}_{\substack{i=|a-a'|+1\\\textrm{by }2}}
  \mathcal{R}^{(2)}(h_{(i,q-\beta,+)};h_{(i,\beta,-)})\\\nonumber
  &\oplus\bigoplus^{a+a'-p-1}_{\substack{\alpha=a+a'-p-1 \textrm{ mod 2}\\ \textrm{by 2}}}\ 
  \bigoplus^{b+b'-q-1}_{\substack{\beta=b+b'-q-1 \textrm{ mod 2}\\ \textrm{by 2}}}\mathcal{R}^{(3)}(h_{(p-\alpha,q-\beta,+)})
\end{align}
Here we used the shorthand
\begin{align}
  \mathcal{R}^{(2)}(h_{(p,j,+)};h_{(0,j,-)})&=\mathcal{W}(h_{(p,j,+)})
  &\mathcal{R}^{(2)}(h_{(i,q,+)};h_{(i,0,-)})&=\mathcal{W}(h_{(i,q,+)})\\\nonumber
  \mathcal{R}^{(3)}(h_{(p,q-b)})&=\mathcal{R}^{(2)}(h_{(p,q-b)})
  &\mathcal{R}^{(3)}(h_{(p-a,q)})&=\mathcal{R}^{(2)}(h_{(p-a,q)})\\\nonumber
  \mathcal{R}^{(3)}(h_{(p,q)})&=\mathcal{W}(h_{(p,q)}).
\end{align}

\section{Characters}
\label{sec:chars}

The characters of the reducible but indecomposable representations can be expanded in terms of \(\chi_{(a,b,\mu)}\),
the characters of the irreducible representations \({\cal W}(h_{(a,b,\mu)})\)
\begin{align*}
  \chi[{\cal W}_{a,b}]&=\chi[{\cal W}_{a,b}^\ast]=\chi_{(a,b,0)}+\chi_{(a,b,+)}\\
  \chi[{\cal R}^{(2)}(h_{(a,b,\pm)};h_{(p-a,b,\mp)})]&=\delta_{+,\pm}\cdot\chi_{(a,b,0)}+2\chi_{(a,b,\pm)}+2\chi_{(p-a,b,\mp)}\\
  \chi[{\cal R}^{(2)}(h_{(a,b,\pm)};h_{(a,q-b,\mp)})]&=\delta_{+,\pm}\cdot\chi_{(a,b,0)}+2\chi_{(a,b,\pm)}+2\chi_{(a,q-b,\mp)}\\
  \chi[{\cal R}^{(2)}(h_{(r,q,\pm)})]&=2\chi_{(r,q,\pm)}+2\chi_{(p-r,q,\mp)}\\
  \chi[{\cal R}^{(2)}(h_{(p,s,\pm)})]&=2\chi_{(p,s,\pm)}+2\chi_{(p,q-s,\mp)}\\
  \chi[{\cal R}^{(3)}(h_{(a,b,+)})]&=2\chi_{(a,b,0)}+4\chi_{(a,b,+)}+4\chi_{(p-a,q-b,+)}+4\chi_{(p-a,b,-)}+4\chi_{(a,q-b,-)}\\
  \chi[{\cal R}^{(3)}(h_{(a,b,-)})]&=2\chi_{(p-a,b,0)}+4\chi_{(p-a,b,+)}+4\chi_{(a,q-b,+)}+4\chi_{(a,b,-)}+4\chi_{(p-a,q-b,-)}.
\end{align*}
We therefore have the following equalities among characters and thus also among classes of the Grothendieck group
\(K_0\)
\begin{align}
  \chi[{\cal R}^{(2)}(h_{(r,q,\pm)})]&=\chi[{\cal R}^{(2)}(h_{(r,q,\mp)})]\\\nonumber
  \chi[{\cal R}^{(2)}(h_{(p,s,\pm)})]&=\chi[{\cal R}^{(2)}(h_{(p,s,\mp)})]\\\nonumber
  \chi[{\cal R}^{(2)}(h_{(a,b,+)};h_{(p-a,b,-)})]&=\chi_{(a,b,0)}+\chi[{\cal R}^{(2)}(h_{(p-a,b,-)};h_{(a,b,+)})]\\\nonumber
  \chi[{\cal R}^{(2)}(h_{(a,b,+)};h_{(a,q-b,-)})]&=\chi_{(a,b,0)}+\chi[{\cal R}^{(2)}(h_{(a,q-b,-)};h_{(a,b,+)})]\\\nonumber
  \chi[{\cal R}^{(3)}(h_{(a,b,+)})]&=\chi[{\cal R}^{(3)}(h_{(p-a,q-b,+)})]=\chi[{\cal R}^{(3)}(h_{(p-a,b,-)})]&\\\nonumber
  &=\chi[{\cal R}^{(3)}(h_{(a,q-b,-)})].
\end{align}

\section{Consistency Checks}
\label{sec:consistency}

The fusion rules \eqref{eq:fusionrules} predict that
\begin{align}
  \mathcal{W}(h_{(1,1,+)})\otimes\mathcal{W}(h_{(1,1,+)})&=\mathfrak{N}_+\circ\mathfrak{N}_+
  (\mathcal{W}_{1,1}\otimes\mathcal{W}_{1,1})\\
  &=\mathfrak{N}_\ast(\mathcal{W}_{1,1})=\mathcal{W}_{1,1}^\ast.
\end{align}
We have checked this for \((p,q)=(2,3),(2,5),(3,4),(3,5)\) using the NGK-algorithm \cite{Nahm:1994by,Gaberdiel:1996kx}
 to level 0.
\(\mathcal{W}(h_{(1,1,+)})\) has weight
\begin{align}
  h_{(1,1,+)}=h_{2p-1,1}=h_{1,2q-1}
\end{align}
and therefore nullvectors at levels \(2p-1\) and \(2q-1\). According to \cite{Benoit:1988} these are given by
\begin{align}
  \mathcal{N}_{2p-1}&=\sum_{k=1}^{2p-1}\!\!\!\!\!\!\!\!\!\!\sum_{\substack{\ell_i\geq1\\\ell_1+\cdots+\ell_k=2p-1}}
  \frac{((2p-1)!)^2(-\frac{q}{p})^{2p-1-k}}{\prod_{i=1}^{k-1}(\ell_1+\cdots+\ell_i)(2p-1-\ell_1-\cdots-\ell_i)}
  L_{-\ell_1}\cdots L_{-\ell_k}\mu,\\
  \mathcal{N}_{2q-1}&=\sum_{k=1}^{2q-1}\!\!\!\!\!\!\!\!\!\!\sum_{\substack{\ell_i\geq1\\\ell_1+\cdots+\ell_k=2q-1}}
  \frac{((2q-1)!)^2(-\frac{p}{q})^{2q-1-k}}{\prod_{i=1}^{k-1}(\ell_1+\cdots+\ell_i)(2q-1-\ell_1-\cdots-\ell_i)}
  L_{-\ell_1}\cdots L_{-\ell_k}\mu,
\end{align}
where \(\mu\) is the highest weight vector of \(\mathcal{W}(h_{(1,1,+)})\).

\begin{itemize}
\item$(p,q)=(2,3),\ h_{(1,1,+)}=2$\\
The level 0 space is spanned by \(\mu\otimes_{\mathbb{C}}\mu,\ (L_{-1}\mu)\otimes_{\mathbb{C}}\mu\) and is
obtained by quotienting out \(\mathcal{N}_3\otimes_{\mathbb{C}}\mu,\ 
(L_{-1}\mathcal{N}_3)\otimes_{\mathbb{C}}\mu,\ (L_{-1}^2\mathcal{N}_3)\otimes_{\mathbb{C}}\mu\)
and \(\mathcal{N}_5\otimes_{\mathbb{C}}\mu\). This leads to the relation
\begin{align}
  (L_{-1}^2\mu)\otimes_{\mathbb{C}}\mu\cong-7(L_{-1}\mu)\otimes_{\mathbb{C}}\mu-8\mu\otimes_{\mathbb{C}}\mu.
\end{align}
The \(L_0\)-action is then given by
\begin{align}
  \Delta_{1,0}(L_0)(\mu\otimes_{\mathbb{C}}\mu)&=(L_{-1}\mu)\otimes_{\mathbb{C}}\mu+4\mu\otimes_{\mathbb{C}}\mu\\\nonumber
  \Delta_{1,0}(L_0)(\mu\otimes_{\mathbb{C}}\mu)&=(L_{-1}\mu)^2\otimes_{\mathbb{C}}\mu+5(L_{-1}\mu)\otimes_{\mathbb{C}}\mu\\
  &=-2(L_{-1}\mu)\otimes_{\mathbb{C}}\mu-8\mu\otimes_{\mathbb{C}}\mu.\nonumber
\end{align}
Thus we can represent it by the matrix
\begin{align}
  L_0=\left(
    \begin{array}{cc}
      4&-8\\
      1&-2
    \end{array}
\right)\qquad\textrm{which is conjugate to}\qquad
\left(
    \begin{array}{cc}
      0&0\\
      0&2
    \end{array}
\right),
\end{align}
which is consistent with \(\mathcal{W}_{1,1}^\ast\).
\item$(p,q)=(2,5),\ h_{(1,1,+)}=4$\\
The level 0 space is spanned by \(\mu\otimes_{\mathbb{C}}\mu,\ (L_{-1}\mu)\otimes_{\mathbb{C}}\mu\) and is
obtained by quotienting out \(\mathcal{N}_3\otimes_{\mathbb{C}}\mu,\dots 
,\ (L_{-1}^6\mathcal{N}_3)\otimes_{\mathbb{C}}\mu\)
and \(\mathcal{N}_5\otimes_{\mathbb{C}}\mu\). This leads to the relation
\begin{align}
  (L_{-1}^2\mu)\otimes_{\mathbb{C}}\mu\cong-13(L_{-1}\mu)\otimes_{\mathbb{C}}\mu-32\mu\otimes_{\mathbb{C}}\mu.
\end{align}
The \(L_0\)-action is then given by
\begin{align}
  \Delta_{1,0}(L_0)(\mu\otimes_{\mathbb{C}}\mu)&=(L_{-1}\mu)\otimes_{\mathbb{C}}\mu+8\mu\otimes_{\mathbb{C}}\mu\\\nonumber
  \Delta_{1,0}(L_0)(\mu\otimes_{\mathbb{C}}\mu)&=(L_{-1}\mu)^2\otimes_{\mathbb{C}}\mu+9(L_{-1}\mu)\otimes_{\mathbb{C}}\mu\\
  &=-4(L_{-1}\mu)\otimes_{\mathbb{C}}\mu-32\mu\otimes_{\mathbb{C}}\mu.\nonumber
\end{align}
Thus we can represent it by the matrix
\begin{align}
  L_0=\left(
    \begin{array}{cc}
      8&-32\\
      1&-4
    \end{array}
\right)\qquad\textrm{which is conjugate to}\qquad
\left(
    \begin{array}{cc}
      0&0\\
      0&4
    \end{array}
\right),
\end{align}
which is consistent with \(\mathcal{W}_{1,1}^\ast\).
\item$(p,q)=(3,4),\ h_{(1,1,+)}=6$\\
The level 0 space is spanned by \(\mu\otimes_{\mathbb{C}}\mu,\ (L_{-1}\mu)\otimes_{\mathbb{C}}\mu\) and is
obtained by quotienting out \(\mathcal{N}_5\otimes_{\mathbb{C}}\mu,\dots 
,\ (L_{-1}^4\mathcal{N}_5)\otimes_{\mathbb{C}}\mu\)
and \(\mathcal{N}_7\otimes_{\mathbb{C}}\mu,\dots,\ (L_{-1}^2\mathcal{N}_7)\otimes_{\mathbb{C}}\mu\). This leads to the relation
\begin{align}
  (L_{-1}^2\mu)\otimes_{\mathbb{C}}\mu\cong-19(L_{-1}\mu)\otimes_{\mathbb{C}}\mu-72\mu\otimes_{\mathbb{C}}\mu.
\end{align}
The \(L_0\)-action is then given by
\begin{align}
  \Delta_{1,0}(L_0)(\mu\otimes_{\mathbb{C}}\mu)&=(L_{-1}\mu)\otimes_{\mathbb{C}}\mu+12\mu\otimes_{\mathbb{C}}\mu\\\nonumber
  \Delta_{1,0}(L_0)(\mu\otimes_{\mathbb{C}}\mu)&=(L_{-1}\mu)^2\otimes_{\mathbb{C}}\mu+13(L_{-1}\mu)\otimes_{\mathbb{C}}\mu\\
  &=-6(L_{-1}\mu)\otimes_{\mathbb{C}}\mu-72\mu\otimes_{\mathbb{C}}\mu.\nonumber
\end{align}
Thus we can represent it by the matrix
\begin{align}
  L_0=\left(
    \begin{array}{cc}
      12&-72\\
      1&-6
    \end{array}
\right)\qquad\textrm{which is conjugate to}\qquad
\left(
    \begin{array}{cc}
      0&0\\
      0&6
    \end{array}
\right),
\end{align}
which is consistent with \(\mathcal{W}_{1,1}^\ast\).
\item$(p,q)=(3,5),\ h_{(1,1,+)}=8$\\
The level 0 space is spanned by \(\mu\otimes_{\mathbb{C}}\mu,\ (L_{-1}\mu)\otimes_{\mathbb{C}}\mu\) and is
obtained by quotienting out \(\mathcal{N}_5\otimes_{\mathbb{C}}\mu,\dots 
,\ (L_{-1}^6\mathcal{N}_5)\otimes_{\mathbb{C}}\mu\)
and \(\mathcal{N}_9\otimes_{\mathbb{C}}\mu,\dots,\ (L_{-1}^2\mathcal{N}_9)\otimes_{\mathbb{C}}\mu\). This leads to the relation
\begin{align}
  (L_{-1}^2\mu)\otimes_{\mathbb{C}}\mu\cong-25(L_{-1}\mu)\otimes_{\mathbb{C}}\mu-128\mu\otimes_{\mathbb{C}}\mu.
\end{align}
The \(L_0\)-action is then given by
\begin{align}
  \Delta_{1,0}(L_0)(\mu\otimes_{\mathbb{C}}\mu)&=(L_{-1}\mu)\otimes_{\mathbb{C}}\mu+16\mu\otimes_{\mathbb{C}}\mu\\\nonumber
  \Delta_{1,0}(L_0)(\mu\otimes_{\mathbb{C}}\mu)&=(L_{-1}\mu)^2\otimes_{\mathbb{C}}\mu+17(L_{-1}\mu)\otimes_{\mathbb{C}}\mu\\
  &=-8(L_{-1}\mu)\otimes_{\mathbb{C}}\mu-128\mu\otimes_{\mathbb{C}}\mu.\nonumber
\end{align}
Thus we can represent it by the matrix
\begin{align}
  L_0=\left(
    \begin{array}{cc}
      16&-128\\
      1&-8
    \end{array}
\right)\qquad\textrm{which is conjugate to}\qquad
\left(
    \begin{array}{cc}
      0&0\\
      0&8
    \end{array}
\right),
\end{align}
which is consistent with \(\mathcal{W}_{1,1}^\ast\).
\end{itemize}



\begin{thebibliography}{10}
\bibitem{Jeng:2006tg}
  M.~Jeng, G.~Piroux and P.~Ruelle,
  {\it Height variables in the abelian sandpile model: scaling fields and correlations},
  \doi{10.1088/1742-5468/2006/10/P10015}{J.\ Stat.\ Mech.\ (2006) P10015}
  \arxiv{cond-mat/0609284}{[0609284 [cond-mat]]}.

\bibitem{Pearce:2006we}
P.A.~Pearce and J.~Rasmussen,
{\it Solvable critical dense polymers},
\doi{10.1088/1742-5468/2007/02/P02015}{J.\ Stat.\ Mech.\ (2007) P02015}
\arxiv{hep-th/0610273}{[0610273 [hep-th]]}.

\bibitem{Read:2007qq}
N.~Read and H.~Saleur,
{\it Associative-algebraic approach to logarithmic conformal field theories},
\doi{10.1016/j.nuclphysb.2007.03.033}{Nucl.\ Phys.\  B {\bf 777} (2007) 316}
\arxiv{hep-th/0701117}{[0701117 [hep-th]]}.

\bibitem{Ruelle:2007kg}
P.~Ruelle,
{\it Wind on the boundary for the Abelian sandpile model},
\doi{10.1088/1742-5468/2007/09/P09013}{J.\ Stat.\ Mech.\ (2007) P09013}
\arxiv{0707.3766}{[0707.3766 [cond-mat.stat-mech]]}.

\bibitem{Mathieu:2007pe}
P.~Mathieu and D.~Ridout,
{\it From percolation to logarithmic conformal field theory},
\doi{10.1016/j.physletb.2007.10.007}{Phys.\ Lett.\  B {\bf 657} (2007) 120}
\arxiv{0708.0802}{[0708.0802 [hep-th]]}.

\bibitem{Rasmussen:2008ii}
  J.~Rasmussen and P.A.~Pearce,
  {\it W-extended fusion algebra of critical percolation},
  \doi{10.1088/1751-8113/41/29/295208}{J.\ Phys.\ A  {\bf 41} (2008) 295208}
  \arxiv{0804.4335}{[0804.4335 [hep-th]]}.

\bibitem{Ridout:2008cv}
  D.~Ridout,
  {\it On the percolation BCFT and the crossing probability of Watts},
  \doi{10.1016/j.nuclphysb.2008.09.038}{Nucl.\ Phys.\  B {\bf 810} (2009) 503}
  \arxiv{0808.3530}{[0808.3530 [hep-th]]}.

\bibitem{SaintAubin:2008}
  Y.~Saint-Aubin, P.A.~Pearce and  J.~Rasmussen,
  {\it Geometric Exponents, SLE and Logarithmic Minimal Models}
  \doi{10.1088/1742-5468/2009/02/P02028}{J.\ Stat.\ Mech.\ (2009) P02028}
  \arxiv{0809.4806}{[0809.4806 [cond-mat. stat-mech]]}

\bibitem{Nigro:2009si}
  A.~Nigro,
  {\it Integrals of motion for critical dense polymers and symplectic fermions}
  \arxiv{0903.5051}{[0903.5051 [hep-th]]}.

\bibitem{Gaberdiel:1996np}
  M.R.~Gaberdiel and H.G.~Kausch,
  {\it A rational logarithmic conformal field theory},
  \doi{10.1016/0370-2693(96)00949-5}{Phys.\ Lett.\  B {\bf 386} (1996) 131--137}
  \arxiv{hep-th/9606050}{[9606050 [hep-th]]}.

\bibitem{Fuchs:2003yu}
  J.~Fuchs, S.~Hwang, A.M.~Semikhatov and I.Y.~Tipunin,
  {\it Nonsemisimple fusion algebras and the Verlinde formula},
  \doi{10.1007/s00220-004-1058-y}{Commun.\ Math.\ Phys.\  {\bf 247} (2004) 713}
  \arxiv{hep-th/0306274}{[0306274 [hep-th]]}.

\bibitem{Carqueville:2005nu}
N.~Carqueville and M.~Flohr,
{\it Nonmeromorphic operator product expansion and $C_2$-cofiniteness for a family of W-algebras},
\doi{10.1088/0305-4470/39/4/015}{J.\ Phys.\ A  {\bf 39} (2006) 951}
\arxiv{math-ph/0508015}{[0508015 [math-ph]]}.

\bibitem{Gaberdiel:2007jv}
M.R.~Gaberdiel and I.~Runkel,
{\it From boundary to bulk in logarithmic CFT},
\doi{10.1088/1751-8113/41/7/075402}{J.\ Phys.\ A  {\bf 41} (2008) 075402}
\arxiv{0707.0388}{[0707.0388 [hep-th]]}.

\bibitem{Adamovic:2007er}
D.~Adamovic and A.~Milas,
{\it On the triplet vertex algebra W(p)},
\doi{10.1016/j.aim.2007.11.012}{Adv.\ Math.\ {\bf 217} (2008) 2664}
\arxiv{0707.1857}{[0707.1857 [math.QA]]}.

\bibitem{nagatomo:2009}
  K.~Nagatomo and A.~Tsuchiya,
  {\it The Triplet Vertex Operator Algebra W(p) and the Restricted Quantum   Group at Root of Unity}
  \arxiv{0902.4607}{[0902.4607 [math.QA]]}.

\bibitem{GabKau96a} 
  M.R.~Gaberdiel and H.G.~Kausch, 
  {\it Indecomposable fusion products}, 
  \doi{10.1016/0550-3213(96)00364-1}{Nucl.\ Phys.\ B {\bf 477} (1996) 293}
  \arxiv{9604026}{[9604026 [hep-th]]}.

\bibitem{GabRunWoo}
  M.R.~Gaberdiel, I.~Runkel and S.~Wood,
  {\it Fusion rules and boundary conditions in the c=0 triplet model}
  \arxiv{0905.0916}{[0905.0916 [hep-th]]}


\bibitem{Rasmussen:2008xi}
J.~Rasmussen,
{\it W-extended logarithmic minimal models},
\doi{10.1016/j.nuclphysb.2008.07.029}{Nucl.\ Phys.\  B {\bf 807} (2009) 495}
\arxiv{0805.2991}{[0805.2991 [hep-th]]}.

\bibitem{Rasmussen:2008ez}
J.~Rasmussen,
{\it Polynomial fusion rings of W-extended logarithmic minimal models},
\doi{10.1063/1.3093265}{J.\ Math.\ Phys.\ {\bf 50} (2009) 043512}
\arxiv{0812.1070}{[0812.1070 [hep-th]]}


\bibitem{Feigin:2006iv}
  B.L.~Feigin, A.M.~Gainutdinov, A.M.~Semikhatov and I.Y.~Tipunin,
  {\it Logarithmic extensions of minimal models: Characters and modular transformations},
  \doi{10.1016/j.nuclphysb.2006.09.019}{Nucl.\ Phys.\  B {\bf 757} (2006) 303}
  \arxiv{hep-th/0606196}{[0606196 [hep-th]]}.

\bibitem{Belavin:1984}
  A.A.~Belavin, A.M.~Polyakov and A.B.~Zamolodchikov,
  {\it Infinite conformal symmetry in two-dimensional quantum field theory},
  \doi{10.1016/0550-3213(84)90052-X}{Nucl.\ Phys.\ B {\bf 241} (1984) 333}

\bibitem{Nahm:1994by}
W.~Nahm,
{\it Quasirational fusion products},
\doi{10.1142/S0217979294001597}{Int.\ J.\ Mod.\ Phys.\  B {\bf 8} (1994) 3693}
\arxiv{hep-th/9402039}{[9402039 [hep-th]]}.

\bibitem{Gaberdiel:1996kx}
M.R.~Gaberdiel and H.G.~Kausch,
{\it Indecomposable fusion products},
\doi{10.1016/0550-3213(96)00364-1}{Nucl.\ Phys.\ B {\bf 477} (1996) 293}
\arxiv{hep-th/9604026}{[9604026 [hep-th]]}.

\bibitem{Quella:2007hr}
T.~Quella and V.~Schomerus,
{\it Free fermion resolution of supergroup WZNW models},
\doi{10.1088/1126-6708/2007/09/085}{JHEP {\bf 0709} (2007) 085}
\arxiv{0706.0744}{[0706.0744 [hep-th]]}.

\bibitem{Runkel:1998pm}
I.~Runkel,
{\it Boundary structure constants for the A-series Virasoro minimal models},
\doi{10.1016/S0550-3213(99)00125-X}{Nucl.\ Phys.\  B {\bf 549} (1999) 563}
\arxiv{hep-th/9811178}{[9811178 [hep-th]]}.

\bibitem{Felder:1999mq}
G.~Felder, J.~Fr\"ohlich, J.~Fuchs and C.~Schweigert,
{\it Correlation functions and boundary conditions in RCFT and three-dimensional topology},
\doi{10.1023/A:1014903315415}{Compos.\ Math.\  {\bf 131} (2002) 189}
\arxiv{hep-th/9912239}{[9912239 [hep-th]]}.
 
\bibitem{tft1}
J.~Fuchs, I.~Runkel and C.~Schweigert,
{\it TFT construction of RCFT correlators. I: Partition functions},
\doi{10.1016/S0550-3213(02)00744-7}{Nucl.\ Phys.\ B {\bf 646} (2002) 353} 
\arxiv{hep-th/0204148}{[0204148 [hep-th]]}.

\bibitem{unique}
J.~Fjelstad, J.~Fuchs, I.~Runkel and C.~Schweigert,
{\it Uniqueness of open/closed rational CFT with given algebra of open states},
Adv.\ Theor.\ Math.\ Phys.\ {\bf 12} (2008) 1283
\arxiv{hep-th/0612306} {[0612306 [hep-th]]}.

\bibitem{Kong:2008ci}
L.~Kong and I.~Runkel,
{\it Cardy algebras and sewing constraints, I},
\doi{doi:10.1016/j.aim.2008.07.004}{Adv.\ Math.\ {\bf 219} (2008) 1548}
\arxiv{0708.1897}{[0708.1897 [math.QA]]}.

\bibitem{Rasmussen:2009}
J.~Rasmussen,
{\it Fusion of irreducible modules in WLM(p,p')}
\arxiv{0906.5414}{[0906.5414 [hep-th]]}

\bibitem{Benoit:1988}
L.~Benoit and Y.~Saint-Aubin,
{\it Degenerate conformal field theories and explicit expressions for some null vectors},
\doi{10.1016/0370-2693(88)91352-4}{Phys.\ Lett.\ B   {\bf 215} (1988) 517--522}

\end{thebibliography}
\end{document}